\documentclass[preprint]{revtex4}
\usepackage{graphicx}
\usepackage{dcolumn}
\usepackage{amsmath}

\begin{document}
\newcommand \bfig {\begin{figure}}
\newcommand \efig {\end{figure}}

\draft
\title{A key to room-temperature ferromagnetism in Fe-doped ZnO: Cu}

\author{
S-J. Han, J. W. Song, C. -H. Yang, S. H. Park, J.-H. Park, and Y.
H. Jeong}\email[electronic mail: ]{yhj@postech.ac.kr}
\affiliation{Department of Physics and electron Spin Science
Center, Pohang University of Science and Technology, Pohang,
790-784, S. Korea}
\author{K. W. Rhie}
\address{Department of Physics, Korea University, Chochiwon 339-700, S. Korea\vskip 1cm}


\begin{abstract}
Successful synthesis of room-temperature ferromagnetic
semiconductors, Zn$_{1-x}$Fe$_{x}$O, is reported. The essential
ingredient in achieving room-temperature ferromagnetism in bulk
Zn$_{1-x}$Fe$_{x}$O was found to be additional Cu doping. A
transition temperature as high as 550 K was obtained in
Zn$_{0.94}$Fe$_{0.05}$Cu$_{0.01}$O; the saturation magnetization
at room temperature reached a value of $0.75\,\mu _{\rm B}$ per
Fe. Large magnetoresistance was also observed below $100\,$K.
\end{abstract}


\maketitle

Diluted magnetic semiconductors (DMSs) have attracted a great deal
of attention because of the possibility of incorporating the
magnetic degrees of freedom in traditional semiconductors
\cite{FurdynaJK_JAP_88,wolf}. DMSs combine their transport and/or
optical properties with magnetism and thereby carry an enormous
potential of opening up a path to entirely new devices. Until
recently, (Ga,Mn)As has been a representative DMS with its
moderately high Curie point (maximum $T_c\,\approx\,$110 K)
\cite{OhnoH_JMMM_99}.
However, an essential task in a device realization of the
potential is to find a DMS with the Curie point above room
temperature
\cite{MedvedkinGA_etal_JJAP_00,MatsumotoY_etal_S_01,reed}.

Recent theoretical works predicted ferromagnetism above room
temperature in a II-VI semiconductor ZnO, normally of n-type, when
doped with magnetic impurities
\cite{DietlT_etal_S_00,SatoK_K-YoshidaH_JJAP_01}. Since ZnO is
optically transparent, ferromagnetic ZnO would be a transparent
magnet as well. Despite of intensive efforts on transition
metal-doped ZnO {\it thin films}, the experimental results did not
converge on a definite conclusion; there are, for instance,
contradicting reports on Co-doped ZnO thin films
\cite{JinZ_etal_APL_01,UedaK_etal_APL_01}. Even the successful
report, in which Co-doped ZnO thin films showed $T_c$ of about 300
K, was attached with reservation that the reproducibility was less
than 10\% \cite{UedaK_etal_APL_01}. Surveying the current
situation with transition metal-doped ZnO, one naturally comes to
a suspicion that the inconsistent results in thin film DMSs,
within a group or among different groups, might be due to
sensitive dependence of thin films on detailed process conditions
such as target qualities, substrates, growing temperatures, oxygen
pressure, etc. These thoughts motivated us to probe {\it bulk}
samples rather than thin films.

In this letter, we concentrate on Fe-doped ZnO bulk samples. As
described below, Fe-doping alone turned out to be not sufficient
for room temperature ferromagnetism in ZnO and the third element
was required. As one member of the transition metal group, Cu
substitution may be considered as magnetic doping if Cu
substitutes for Zn$^{2+}$ as Cu$^{2+}$. Cu-doping, however, did
not induce a significant change in the magnetic property of ZnO
films \cite{JinZ_etal_APL_01}. On the other hand, Cu may be used
as an additional p-type dopant into naturally n-type ZnO samples
\cite{JunST_ChoiGM_JACerS_98}. This new idea of additional
Cu-doping in Zn$_{1-x}$Fe$_x$O was highly successful and led us to
a room temperature ferromagnetic DMS. In view of the inconsistent
film results, reproducibility was ascertained by measuring several
samples synthesized by the same procedures.

Polycrystalline samples were fabricated with the standard solid
state reaction method in Ar filled quartz tubes.
High purity
ZnO(99.99+\%), FeO(99.9+\%), and CuO(99.99+\%) powders were mixed
thoroughly and processed at 1170 K for 24 hours. The single phase
nature of samples was checked by $\theta - 2\theta$ x-ray
diffraction (XRD) using a Cu $K_{\alpha}$ source; substantial
amount of exposure time was allowed in the XRD scans to check even
a minute amount of a secondary phase. Magnetization, resistivity,
Hall coefficient, and thermopower were measured by employing
equipments manufactured by Quantum Design (MPMS and PPMS). The
SQUID magnetometer (MPMS) was equipped with a high temperature
oven facility. X-ray absorption spectroscopy (XAS) was carried out
at the Dragon beamline of the Synchrotron Radiation Research
Center in Taiwan.

Fig. 1(a) shows a typical powder XRD pattern of Zn$_{1-x}$Fe$_x$O
for $ x = 0.07 $. All the peaks belong to the hexagonal lattice of
ZnO, and no indication of a secondary phase is found. A shift of
XRD peak positions related to lattice spacing changes was clearly
observed when the concentration of Fe was varied; the data
refinement revealed that the shift was caused by a variation of
lattice spacing $a$ as displayed in Fig. 1(b). The linear
expansion of the $a$-axis lattice spacing with increasing $x$
indicates that doped Fe atoms substitute for Zn atoms in the
lattice up to $ x = 0.07 $ under the current processing condition.
The incorporation of Fe atoms into the lattice was also evidenced
from the XAS measurements which yielded the oxidation state of Fe
to be mostly Fe$^{2+}$. On further doping of Fe above
$x\,\approx\,0.1$, the system enters a coexistence region of the
dominant hexagonal phase (Fe-doped ZnO) and a minor cubic  phase
(FeO). It should be noted here that the Fe solubility of bulk ZnO
found in this work is higher than that of thin films
\cite{JinZ_etal_APL_01}. The low solubility of the films seems to
be related to the fact that it is the $c$-axis lattice constant,
not $a$-axis which is constrained by a substrate, that varies as
the contents of Fe changes. This is a manifestation of the
difference between bulk and thin films, which will be discussed
elsewhere \cite{song}.

We now turn to the magnetic properties of Zn$_{1-x}$Fe$_x$O. When
measured at room temperature, a sample with $x$ = 0.05 showed a
maximum saturation  magnetization ($M_s$). However, the measured
$M_s$ of Zn$_{0.95}$Fe$_{0.05}$O (0.025$\mu_B$/Fe, $\mu_B$ = Bohr
magneton) was far too small to be considered as a room temperature
ferromagnet. The situation is evident in Fig. 2(a), showing the
magnetization$\!-\!$field ($M\!-\!H$) curve of
Zn$_{0.95}$Fe$_{0.05}$O sample at 300$\,$K. We then attempted
additional Cu-doping into Zn$_{0.95}$Fe$_{0.05}$O as a means of
rendering respectable ferromagnetism to the system. According to
Dietl {\it et al.}, hole doping is most effective in achieving
carrier-mediated ferromagnetism in semiconductors
\cite{DietlT_etal_S_00}. Jun and Choi showed that introduction of
CuO into n-type ZnO increases resistivity up to 1\% mixture
\cite{JunST_ChoiGM_JACerS_98}, which suggests that Cu can play the
role of a p-type dopant. Cu-doping into Zn$_{0.95}$Fe$_{0.05}$O
via the standard ceramic method neither caused any structure
change nor induced a secondary phase up to 1\% incorporation. A
small amount of additional Cu-doping, however, brought about
drastic changes in $M$ as illustrated in Fig. 2(b), (c) and (d).
The magnetization is greatly enhanced with Cu-doping, so that
$M_s$ at room temperature of the sample with 1\% Cu
(0.75$\mu_B$/Fe) becomes 30 times larger than that of the sample
without Cu. The fact that $M_s$ increases systematically only with
Cu contents virtually eliminates the possibility of Fe (or iron
oxide) clusters being responsible for magnetization. The
saturation occurs at 2$\,$kOe, while the coercive field is about
20$\,$Oe at room temperature. The small coercivity of a
polycrystalline sample indicates the intrinsically soft nature of
this material.

Now that we have established ferromagnetism in
Zn$_{0.95-y}$Fe$_{0.05}$Cu$_{y}$O at room temperature, its Curie
temperature is of interest. In order to determine $T_c$ of
Zn$_{0.94}$Fe$_{0.05}$Cu$_{0.01}$O, which shows the largest $M_s$
at 300 K in Fig. 2 (c), its $M$ was measured as a function of
temperature at a small field (500 Oe). The measured $M$, presented
in Fig. 3, clearly shows that a transition to a paramagnetic state
occurs near 550 K. It is noted that the transition is rather sharp
and appears to deviate from a mean field description. The
transition temperature of 550 K is considerably higher than the
Curie temperature of most DMSs. More importantly, it is high
enough for the purpose of device applications at room temperature.

In addition to ferromagnetism, transport properties are of
exceeding importance in DMS. Fig. 4(a) is the plot of the
resistivity of Zn$_{0.94}$Fe$_{0.05}$Cu$_{0.01}$O at zero field as
a function of temperature. The resistivity displays a typical
semiconducting behavior. At room temperature the resistivity has a
value of 0.1 $\Omega$cm. Magnetoresistance (MR
$\equiv\,(\rho(H)-\rho(0))/\rho(0)$) was also measured at various
temperatures. The measured MR of a bulk sample is very large and
positive as shown in Fig. 4(b). However, the MR of
Zn$_{0.94}$Fe$_{0.05}$Cu$_{0.01}$O becomes insignificant above 100
K, even if the system is still ferromagnetic. At present, we have
no understanding on this peculiar behavior.

In order to elucidate the origin of ferromagnetism, we also
measured the thermopower and the Hall coefficient of
Zn$_{0.95}$Fe$_{0.05}$O and Zn$_{0.94}$Fe$_{0.05}$Cu$_{0.01}$O.
Both samples yielded negative values for thermopower as well as
Hall coefficient, and no anomalous Hall effect was observed. These
results, first of all, indicate that electrons are major carriers
in the samples. The number of carriers at room temperature
estimated from the Hall coefficients are 4.2x$10^{17}$/cm$^3$ and
5.0x$10^{17}$/cm$^3$ for Zn$_{0.94}$Fe$_{0.05}$Cu$_{0.01}$O and
Zn$_{0.95}$Fe$_{0.05}$O, respectively. XAS measurements also
showed that the valence state of Cu in
Zn$_{0.94}$Fe$_{0.05}$Cu$_{0.01}$O is Cu$^{1+}$, rather than
Cu$^{2+}$. Thus, Cu ions play the role of acceptors and reduces
the number of electron carriers, although they are not able to
change the n-type character itself. It is then evident that the
ferromagnetism in Zn$_{0.94}$Fe$_{0.05}$Cu$_{0.01}$O (which
appears with reduction in the number of electrons) is not mediated
by electrons. As Sato {\it et al.} suggested, it is very likely
that a 3$d$ band is formed by doped Fe atoms, and ferromagnetism
occurs within this band via double exchange
\cite{SatoK_K-YoshidaH_JJAP_01}. The lack of anomalous Hall effect
is probably due to a weak coupling of the conduction electrons,
which dominate transport, to magnetic degrees of freedom.

In conclusion, we achieved  ferromagnetism in Zn$_{1-x}$Fe$_x$O by
Cu-doping, and the Curie temperature of 550 K and the saturation
magnetization of 0.75$\mu_B$/Fe were obtained for
Zn$_{0.94}$Fe$_{0.05}$Cu$_{0.01}$O.


This work was supported by the SRC program of KOSEF and the BK21
program of MOE. J.-H. Park thanks Drs. H.-J. Lin and C.T. Chen of
the SRRC in Taiwan for the beam time.

\newpage

\newpage

{\bf\centering Figure Captions}\\

\noindent FIG. 1: (a) Intensity of X-ray diffraction vs $2\theta$,
from a polycrystalline Zn$_{1-x}$Fe$_x$O sample with $x = 0.07$.
All the peaks belong to the hexagonal structure of ZnO.
(b) Variation of the lattice constant $a$ vs Fe concentration $x$.\\

\noindent FIG. 2: Room temperature $M-H$ curves of bulk
Zn$_{0.95-y}$Fe$_{0.05}$Cu$_y$O: (a) $y$ = 0, (b) 0.002, (c)
0.005, (d) 0.01. The vertical scale (magnetic moment/Fe) was
converted from measured values,
assuming homogeneous magnetic states.\\

\noindent FIG. 3: The magnetization of
Zn$_{0.94}$Fe$_{0.05}$Cu$_{0.01}$O is plotted as a function of
temperature. The strength of the applied field was 500 Oe.\\

\noindent FIG. 4: Transport properties of
Zn$_{0.94}$Fe$_{0.05}$Cu$_{0.01}$O: (a) The resistivity vs
temperature, (b) Magnetoresistance (MR) at various temperatures.
MR is defined as $(\rho(H)-\rho(0))/\rho(0)$. At 100 K, the MR is
insignificant.

\newpage

\bfig \includegraphics[width=17cm]{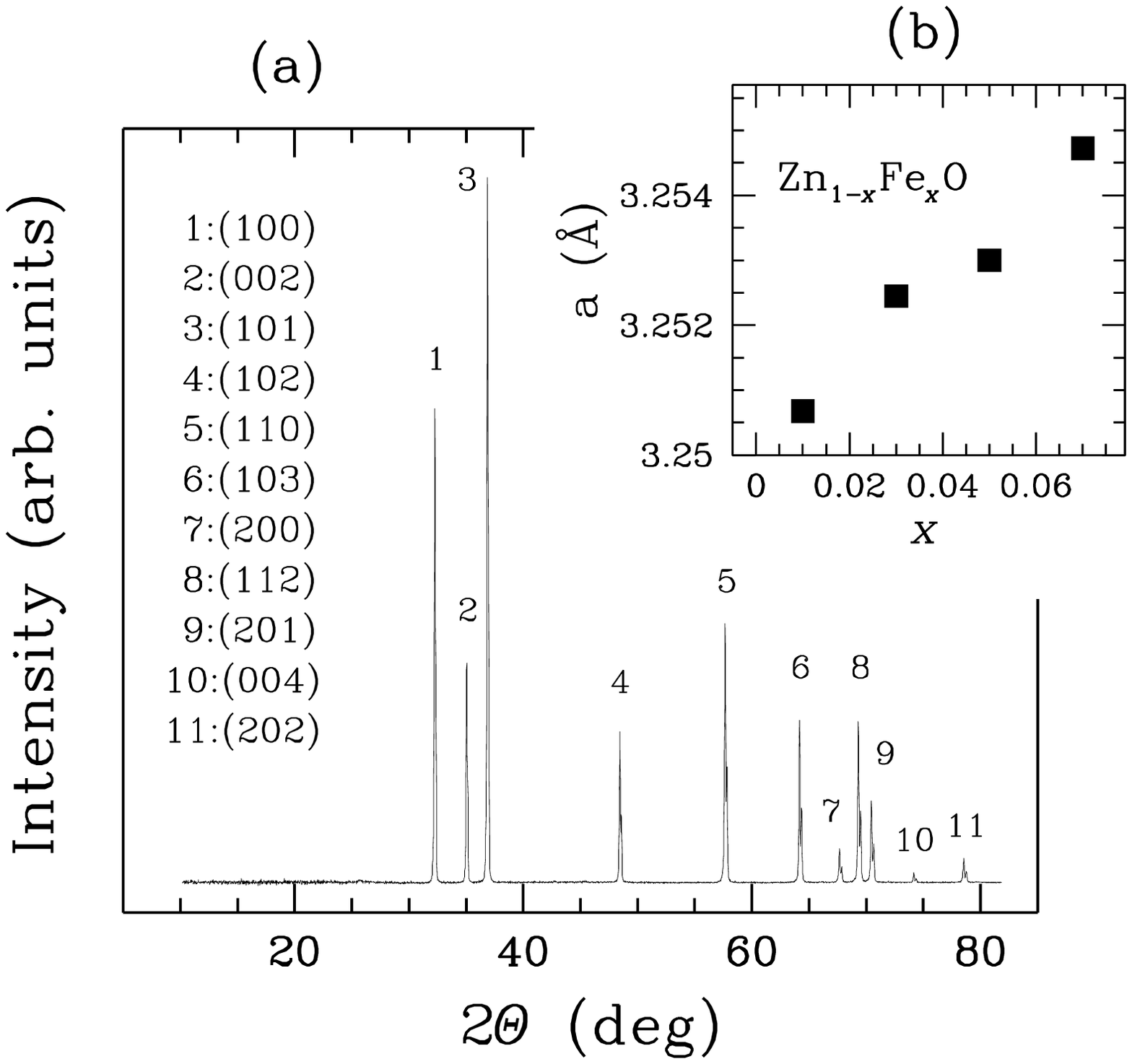} \vskip 3cm\caption{
Han et al.} \efig

\bfig \includegraphics[width=17cm]{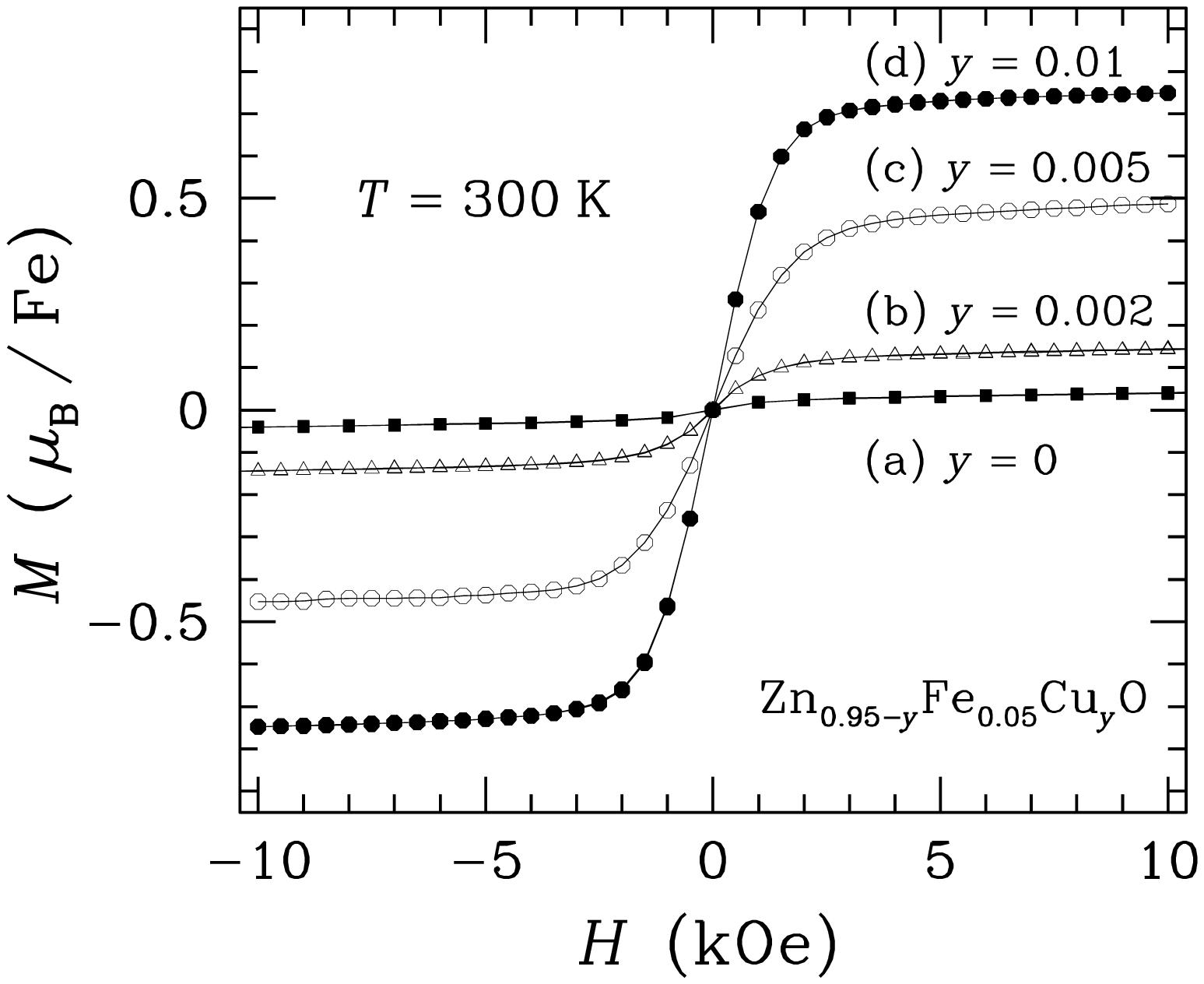}  \vskip
3cm\caption{Han et al. } \efig

\bfig \includegraphics[width=17cm]{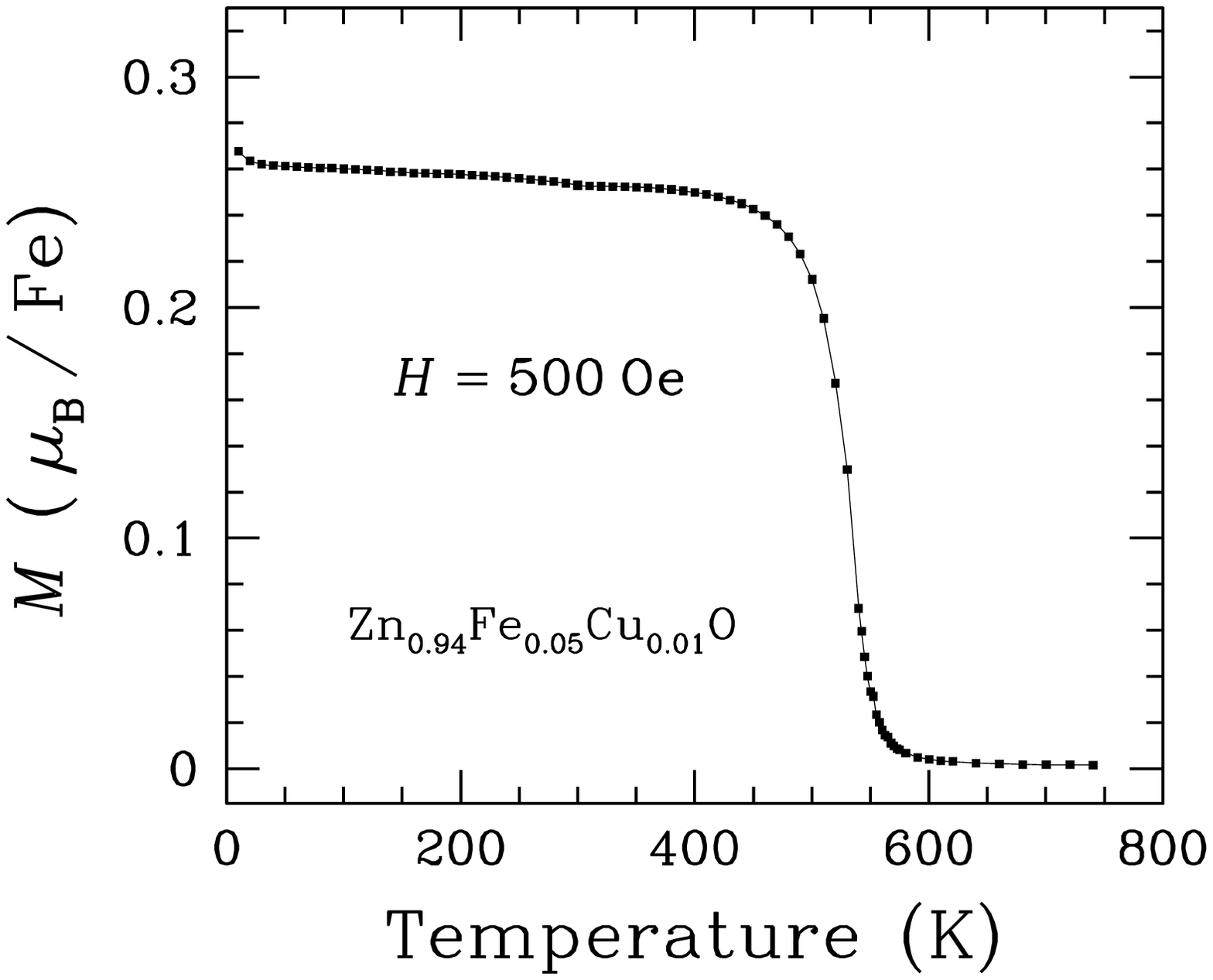}  \vskip
3cm\caption{Han et al. }\efig

\bfig \includegraphics[width=17cm]{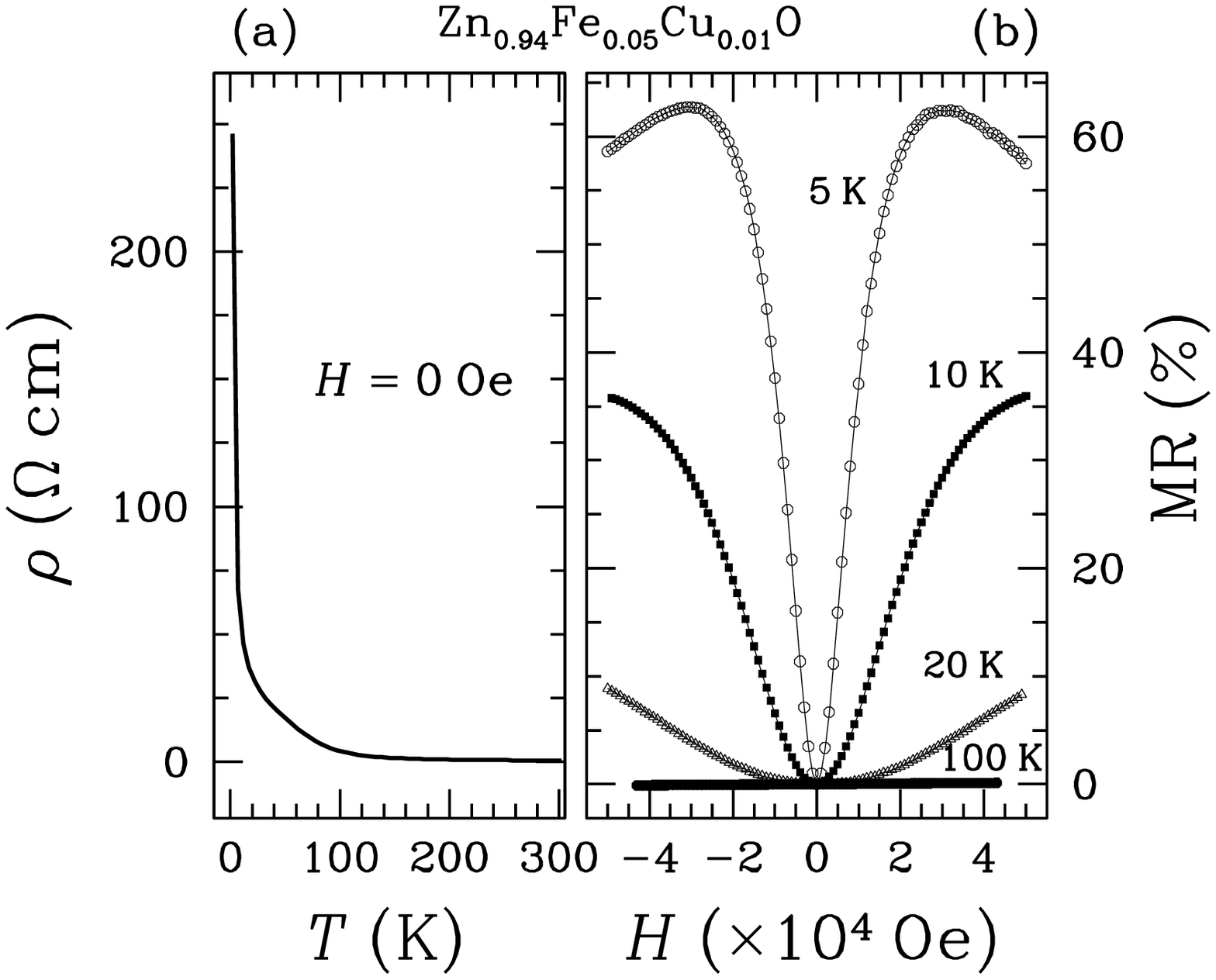}  \vskip
3cm\caption{Han et al. }\efig

\end{document}